\documentclass{emulateapj}
\pdfoutput=1

\def\rj{$R_{\rm J}\ $}
\def\etal{{et~al.\,}}

\def\mp{M$_{\rm p}$}
\def\rp{R$_{\rm p}\,$}

\def\mic{$\mu$m$\,$}

\def\sles{\lower2pt\hbox{$\buildrel {\scriptstyle <}
   \over {\scriptstyle\sim}$}}
\def\sgreat{\lower2pt\hbox{$\buildrel {\scriptstyle >}
   \over {\scriptstyle\sim}$}}

\slugcomment{Accepted to Ap.J. June 14, 2010}

\begin{document}

\title{Photometric and Spectral Signatures of 3D Models of Transiting Giant Exoplanets}  

\author{A. Burrows\altaffilmark{1,2}, E. Rauscher\altaffilmark{3,2}, D.S. Spiegel\altaffilmark{1,2}, \& K. Menou\altaffilmark{3,2}} 

\altaffiltext{1}{Department of Astrophysical Sciences, 
                 Peyton Hall, Princeton University, Princeton, NJ 08544; burrows@astro.princeton.edu, dsp@astro.princeton.edu}

\altaffiltext{2}{Kavli Institute for Theoretical Physics, University of California, Santa Barbara, CA 93106}

\altaffiltext{3}{Department of Astronomy, Columbia University, 550 West 120th Street, New York, NY 10027; emily@astro.columbia.edu, kristen@astro.columbia.edu}

\begin{abstract}

Using a 3D GCM, we create dynamical model atmospheres of
a representative transiting giant exoplanet, HD 209458b. We post-process these
atmospheres with an opacity code to obtain transit radius
spectra during the primary transit. Using a spectral atmosphere code, we
integrate over the face of the planet seen by an observer at various orbital phases
and calculate light curves as a function of wavelength and for
different photometric bands. The products of this study are
generic predictions for the phase variations of a zero-eccentricity giant planet's
transit spectrum and of its light curves. We find that for these
models the temporal variations in all quantities and the ingress/egress
contrasts in the transit radii are small ($< 1.0$\%). Moreover, we determine
that the day/night contrasts and phase shifts of the brightness peaks relative to the
ephemeris are functions of photometric band.  The $J$, $H$, and $K$ bands are shifted most,
while the IRAC bands are shifted least.  Therefore, we verify that the magnitude
of the downwind shift in the planetary ``hot spot" due to equatorial winds is
strongly wavelength-dependent.  The phase and wavelength
dependence of light curves, and the associated day/night contrasts, can be used to
constrain the circulation regime of irradiated giant planets and to probe different pressure
levels of a hot Jupiter atmosphere.  We posit that though our calculations focus on models of
HD 209458b similar calculations for other transiting hot Jupiters in low-eccentricity
orbits should yield transit spectra and light curves of a similar character.

\end{abstract}


\section{Introduction}
\label{intro}

Circa April 2010, more than 70 exoplanets had been discovered  
transiting their primary stars\footnote{see J. Schneider's
Extrasolar Planet Encyclopaedia at http://exoplanet.eu, the Geneva Search Programme at
http://exoplanets.eu, and the Carnegie/California compilation at http://exoplanets.org}.  Transits are useful in the characterization
of exoplanets because they break the the mass/orbital-inclination degeneracy 
of radial-velocity measurements.  Moreover, a planet's radius can be directly measured,
yielding mass (\mp) and radius (\rp) correlations with which theorists can
extract useful information about bulk composition and structure and can attempt to fit 
radius evolution models (Guillot et al. 1996; Burrows et al. 2000; Brown et al. 2001; 
Hubbard et al. 2001; Baraffe et al. 2003; Chabrier et al. 2004; Charbonneau et al. 2007). 

However, since atmospheric opacity is a function of wavelength, a planet's transit 
radius is also a function of wavelength (Seager \& Sasselov 2000; Brown 2001;
Spiegel et al. 2007; Fortney et al. 2003,2010).  The variation with wavelength in the measured radius 
(actually an impact parameter) provides an ersatz spectrum directly related to 
the planet's atmospheric composition near the terminator.  This radius spectrum,
unlike the planet's spectrum at secondary eclipse, is more dependent
upon composition than upon the temperature profile,
and so provides complementary information to a planet's own direct emissions.  The latter can
be measured during secondary eclipse, but also during the traverse by the planet 
of its orbit as it traces out its phase light curve.  

The transit radius spectrum and the planet emission spectrum as a function of phase
can, therefore, together help constrain planet properties (Fortney et al. 2006,2010).  However, the stellar irradiation   
of such an exoplanet severely breaks what would otherwise be quasi-spherical symmetry, 
producing large day-night contrasts in thermal structure, zonal flows and banding, and
violent atmospheric dynamics (Showman \& Guillot 2002; Guillot \& Showman 2002; 
Cho et al. 2003,2008; Burkert et al. 2005; Cooper \& Showman 2005,2006; 
Showman et al. 2008,2009; Showman, Cho, \& Menou 2010; Dobbs-Dixon \& Lin 2008; Dobbs-Dixon, Cumming, \& Lin 2010; Rauscher et 
al. 2007,2008; Menou \& Rauscher 2009; Rauscher \& Menou 2010).  Importantly, such behavior may have photometric and spectral 
signatures. It is the exploration of such signatures that motivates this paper.
The state-of-the-art in the 3D modeling of exoplanet atmospheres is still evolving and 
has not yet reached a level of maturity where detailed predictions for each known exoplanet
are robust.  Hence, in this paper we focus on a few generic ideas and conjectures that
emerge from our modeling efforts.  The procedure we have pursued is the following:  First,
using a 3D general circulation model (GCM) we derive dynamic model atmospheres at various epochs after they have achieved a steady state.
Second, we post-process these 3D model atmospheres with a spectral atmosphere code to obtain transit 
radius spectra at various epochs during ingress, egress, and total transit.  We assume 
equilibrium molecular compositions and solar metallicity.  Third, using the same 3D model
atmosphere, we integrate over the ``visible" disk and calculate phase light curves as 
a function of wavelength and for various standard photometric bands.  The results are
generic predictions which, though not expected to be quantitatively precise and constraining
on any particular giant exoplanet, nevertheless contain qualitatively interesting features
that should inform future measurements.  For specificity, we focus on two 
models of HD 209458b (Charbonneau et al. 2000; Henry et al. 2000), one with and one 
without an ``extra absorber" and a thermal inversion (Hubeny, Burrows, \& Sudarsky 2003; 
Burrows et al. 2007; Fortney et sl. 2008; Spiegel et al. 2009; Knutson et al. 2008), but suggest our qualitative 
results are generic beyond this planet and this modeling paradigm (see also Fortney et al. 2006,2010).

In \S\ref{techniques}, we summarize our methodology and techniques.  Then, in \S\ref{describe} we describe
the 3D models and various of their salient characteristics. We go on in \S\ref{wave_rad} to present
our results for the wavelength-dependent transit radius and ingress-egress asymmetries and in
\S\ref{phase} we turn to a full discussion of the derived light curves.  This section contains not only 
wavelength-dependent planet-star flux ratios as a function of phase, but the phase variation
of several photometric planet band fluxes.  In \S\ref{conclusions}, we summarize our general conclusions.

\section{Methodology}
\label{techniques}

\subsection{3D GCM Models}
\label{gcm}

\begin{figure}
\includegraphics[height=.60\textheight,angle=0]{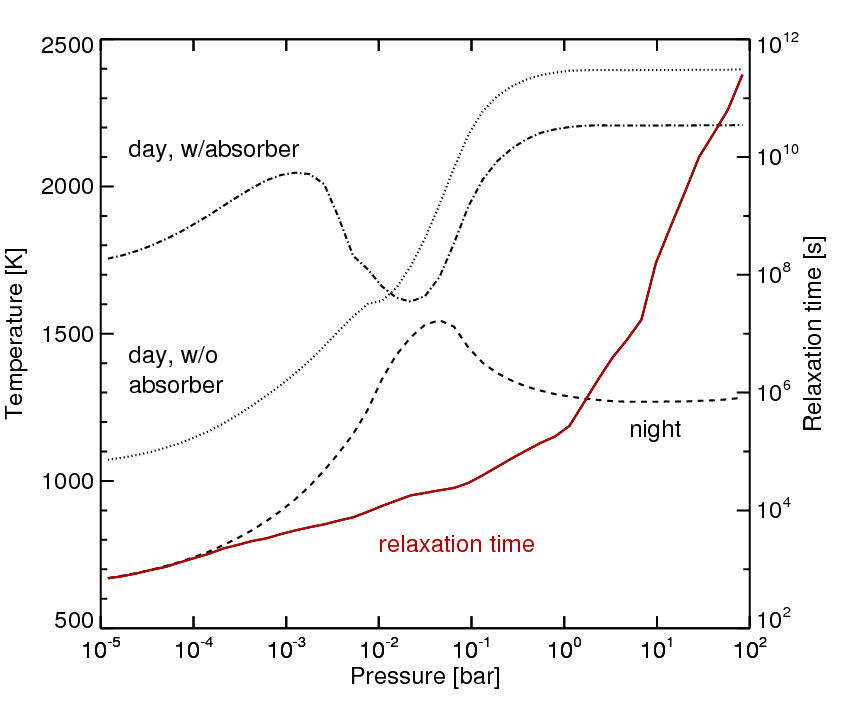}
\caption{The relaxation temperature-pressure profiles used for the forcing.  The night side profile (dashed) is the same for both models, with separate substellar     
point profiles for the models with (dot-dashed) and without (dotted) an extra absorber high in the atmosphere.  The solid red line shows the relaxation       
times used at each pressure level, for both models.  See text for a description of the forcing scheme.} 
\label{fig1}
\end{figure}

We use the same modified version of the University of Reading's
Intermediate General Circulation Model (Hoskins \& Simmons 1975) as
presented in Menou \& Rauscher (2009) and Rauscher \& Menou (2010),
where a detailed description of the numerical implementation can be
found.  We use a horizontal spectral resolution of T31, roughly
equivalent to a 4$^{\circ}$ resolution on the sphere, and 45 vertical
levels which are logarithmically spaced in pressure from 100 bar to 10
microbar.  We employ a simplified Newtonian relaxation scheme for the
radiative forcing, so that everywhere the atmosphere heats or cools
toward a prescribed three-dimensional equilibrium temperature profile,
on some representative radiative timescale (Iro, B\'{e}zard, \&
Guillot 2005).  
The equilibrium temperature profiles are chosen to have a horizontal
dependence such that on the dayside the fourth power of the equilibrium
temperature goes linearly with the cosine of the angle from the substellar
point and on the nightside it is constant.  The amplitude of the
temperature difference between the substellar point and the nightside
is taken from 1D radiative transfer models (Burrows et al. 2008; see
Fig. \ref{fig1}), as are the dependences with pressure and depth of
the equilibrium temperatures needed in the Newtonian scheme.

Our choice for the angular dependence of the dayside equilibrium
temperatures is loosely motivated by the fact that the equilibrium
black body temperature ($T_{eq}$) of an irradiated surface at a
slant angle is determined by setting the incident flux on the angled
surface equal to the emitted flux and by the $\sigma T_{eq}^4$
dependence of the latter.  We scale the entire vertical profile
of internal equilibrium temperatures to which the Newtonian cooling
scheme is dynamically driving the actual temperature by this same
angular factor.  In this way, the expectation that the annuli
on the planet away from the substellar point will be heated less
by the star than the gas at the substellar point is realized.
For the night side, which by definition is not irradiated by the
star, we employ the simplifying assumption that in equilibrium
without night-side dynamics the planet would resemble more an isolated
brown dwarf with uniform emission and profiles. Note that these
assumptions about the solid-angle distribution of the equilibrium
temperature profiles still allow the 3D GCM dynamics to redistribute
heat in all regions of the planet's atmosphere and that the actual
temperatures perforce deviate from these equilibrium temperatures
due to such advection, sometimes to a significant degree. However,
clearly our ansatz could be improved by doing the 3D radiative
transfer that the problem will eventually demand.  For now, we
believe our approach allows us to capture the essence, if not the
detail, of the thermal character of the dynamical atmospheres we
are studying.

Thus, the only difference between the inputs for our two
3D GCM models of HD 209458b is the different 1D profiles of
the substellar equilibrium temperatures, where one was
calculated including an extra absorber high in the atmosphere and one
was calculated without.  The radiative timescales are constant on a
given pressure level and they increase with pressure at depth.  They
are taken from the profile in Iro et al.  (2005), except for the
deepest model levels. Figure \ref{fig1} provides a detailed profile of
the radiative times adopted.  All other planetary and
numerical parameters match those described in Rauscher \&
Menou (2010).  Each model was run for 500 planetary days ($\equiv$ orbits),
after which a statistically steady state has been reached for
all levels at pressures less than 1 bar.  Levels deeper than this
are not directly probed by the types of observations considered here.
To summarize, the key differences between the present models and those
in Rauscher \& Menou (2010) are (i) the different relaxation
temperature profiles used and (ii) the greater vertical extent of
pressures modeled, up to 10 microbar in the present study.

\subsection{Method for Deriving Transit Curves and Their Wavelength Dependence}
\label{transit_method}

To calculate the spectrum of the transit radius, we create a data cube in latitude, 
longitude, and altitude of the temperatures, densities, and pressures of 
one of the 3D GCM models described in \S\ref{gcm}.  At a given wavelength, using 
the snapshot at day 500 in a 500-day model, we cast 200 rays through an annulus
at a given impact parameter and calculate the optical depth ($\tau_{\lambda}$) along the chord. 
The miniscule index of refraction effects (Fortney et al. 2003) are ignored. We use the opacity 
database described in Sharp \& Burrows (2007) and equilibrium chemical 
abundances taken from Burrows \& Sharp (1999) and Burrows et al. (2001).  Solar abundances were 
assumed, and we have used the inclination angle of 86.6$^{\circ}$ suitable for HD 209458b.  
The differenital area of the annulus associated with a given ray is then weighted by the 
quantity $1 - e^{-{\tau}}$. A new impact parameter is then chosen and the process
is repeated.  By this means, the total effective area of the atmosphere at a given wavelength
is obtained.  It is this effective area that determines the magnitude of the occultation
by the planet of the star due to the atmosphere.  Since there is no core nor structural
information in the atmospheric GCM, we assume that the baseline planet radius
is the radius measured for HD 209458b in the optical by Knutson et al. 2007 (1.32 \rj)
and shift the differential area for a given wavelength calculated as described above 
by the average of the corresponding calculation in the 0.5 to 0.7 \mic wavelength region, 
rendering the average differential area in the 0.5 to 0.7 \mic wavelength region zero.
Other procedures can be followed, but all of them are equivalent to first order
in the (small) differential area, where only the first-order term is relevant.  
When the transit is partial (during ingress and egress), we use the formalism of 
Mandel \& Agol (2002) to determine planet-star intercept angles as a function of 
orbital phase, between which the calculations of the total relevant area of the atmosphere are performed.  
Since the planet is assumed to be in synchronous rotation, it will rotate with the orbit during transit. 
This effect, which can amount to a 10-20$^{\circ}$ turn and which slightly 
changes the regions of the atmosphere intercepted by the cast rays included in 
the optical depth calculation, is incorporated in the formalism.  We perform 
such calculations for 2000 frequencies from 0.4 \mic to 30 \mic and 200 phase angles from just before 
ingress to just after egress. 

Though we performed some transit calculations at other days (e.g., days 451, 476, 496) in the 500-day runs,
we found little variation (less than 0.5\%) in the results.  Hence, we focused
on only the day-500 3D GCM models for the two input models, one with an extra absorber 
and one without, as described in \S\ref{gcm}. Note that such a small temporal variation justifies 
the use of a single snapshot for both these trasnit radius calculations and the phase light curves calculations described below.

\subsection{Spectral Post-Processing to Derive Wavelength-Dependent Light Curves}
\label{spectral_method}

To obtain the total planetary photon fluxes as a function of wavelength, we use the code 
COOLTLUSTY (Hubeny 1988; Hubeny \& Lanz 1995; Burrows, Sudarksy, \& Hubeny 2003; Burrows et al. 2008).
This code incorporates the same composition and opacity algorithms we employ to perform the transit
spectrum calculations described in \S\ref{transit_method}. First, we divide the surface seen by the observer
at a given phase angle into about 2000 patches. Then, we cast a ray from the observer to the patch and
into the planet at the associated slant angle.  The densities, temperatures, and pressures at that slant angle
are then used to calculate the instantaneous radiation field emerging from the patch (without allowing 
the code to seek radiative equilibrium and ignoring the time-dependent term in the transport equation).  
That intensity is then multiplied by the projected area of the patch and all the patches 
seen by the observer at that phase angle were added.  This is done for 1000 frequency points from 
0.4 \mic to 30 \mic.  We also calculated the associated photometric colors in 24 bands, 9 of
which we report here.  The stellar flux is multiplied by the angle of incidence on
the patch with respect to the star and a Kurucz (1994) model for HD 209458 is employed.
The stellar flux is necessary to handle the Rayleigh scattered component and to obtain
the planet/star flux ratio.  We remind the reader that the 3D GCM model (Rauscher \& Menou 2010) is calculated using
Newtonian cooling and not with radiative transfer, so there are inconsistencies in our general approach.
However, our focus here is on the character of generic results (for instance for different 
photometric bands and given general characteristics of the 3D flow), and not on specific 
predictions for HD 209458b.

\section{Description of the 3D GCM Models}
\label{describe}

Although the relaxation temperature profiles are very different for
our two atmospheric models with and without a high-altitude absorber
(Fig. \ref{fig1}), we nevertheless find that these models have rather
similar flow patterns.  This is perhaps even more striking since
these circulation patterns are also qualitatively similar to that
described in Rauscher \& Menou (2010), which has a rather different
temperature relaxation profile.  This suggests that the circulation
pattern studied here is more strongly governed by the choice of 
the $T^4$-behaves-linearly-with-$\cos \theta$ horizontal dependence than by the amplitude or
the detailed vertical structure of the relaxation temperature profile
adopted.  This supports the notion that the main qualitative results
shown in the present study may be extended to other close-in giant
planets.

\begin{figure}
\includegraphics[height=.38\textheight,angle=0]{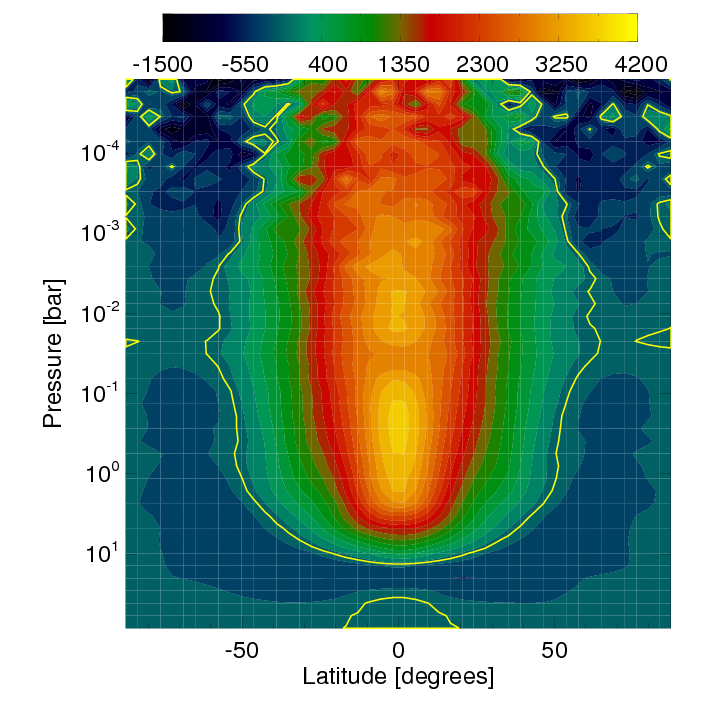}
\includegraphics[height=.38\textheight,angle=0]{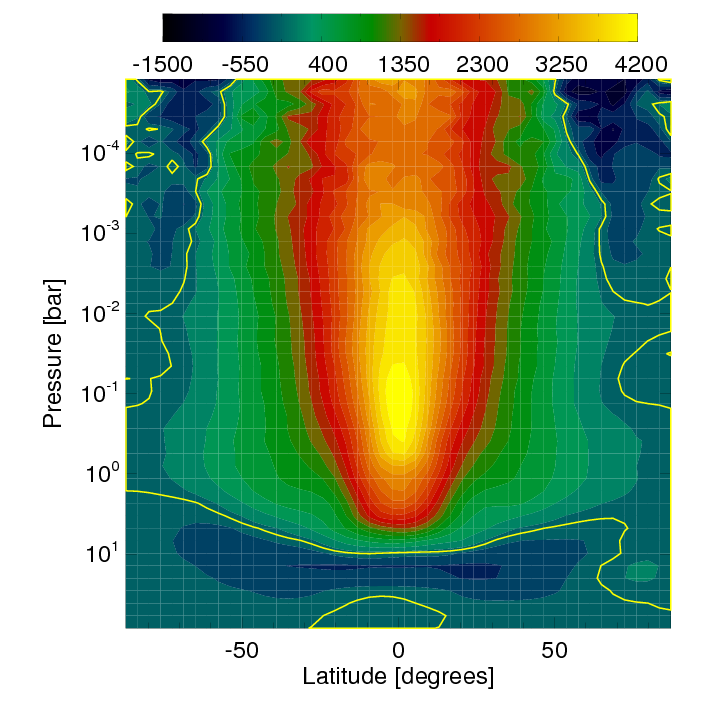}
\caption{The zonal average (i.e., on a latitude circle) of the zonal (east-west) wind in m s$^{-1}$, as a function of the latitude and pressure level in the atmosphere,      
for the model with an absorber (left) and the one without (right).  The solid yellow lines separate regions of postitive (eastward) flow from negative    
pattern is qualitatively similar in both models.  (Also compare to Figure 3 of Rauscher \& Menou 2010.)} 
\label{fig2}
\end{figure}

While details differ, key flow features are common to both model
atmospheres: transonic wind speeds, with peak wind speeds of 10 and 20
km s$^{-1}$ high up in the models with and without an extra absorber,
respectively; a super-rotating (eastward) equatorial jet that extends
across many pressure scale heights (see Fig. \ref{fig2}); 
a pronounced chevron-shaped shock-like feature at a longitude of
$\sim$135$^{\circ}$ at intermediate pressure levels; and large stationary
vortices in each hemisphere that extend almost from the equator to the
poles and are centered at a longitude of $\sim$-120$^{\circ}$ 
(Rauscher \& Menou 2010). High in the atmosphere, where heat advection is not
dominant, the day-night temperature difference is greater in the model
run with an extra absorber because of the stronger forcing profile
adopted. In both models, the horizontal temperature field is close to
equilibrium, with a hot dayside and a cold nightside at pressures
less than 1 mbar. The hottest atmospheric region is advected eastward
by $\sim$45$^{\circ}$ at the 100 mbar level and by $\sim$90$^{\circ}$ at the 1 bar
level, while the temperature field is fairly well longitudinally
homogenized at deeper pressure levels.

The general trend in these models is to have the hottest atmospheric
regions advected farther away from the substellar point at deeper
pressure levels, with potentially observable consequences as discussed
in \S\ref{func_band}.  For the model with the extra absorber, however, there is an
additional complication.  Although both models show localized heating
where a shock-like feature develops in the flow, this heating is
significantly more pronounced in the model with absorber. We attribute
this difference to faster winds driven by a stronger day-night forcing
in the model with absorber.  This, rather than direct heat advection
from the dayside, determines the location of the hottest atmospheric
region at pressure levels 15-45 mbar in the model with absorber. As a
result, the peak temperature region at these levels does not simply
transition away from the substellar point gradually with depth, but it
is significantly affected by a systematic shift to $\sim$135$^{\circ}$, 
where the shock-like feature is located.\footnote{It is worth
recalling that there are modeling concerns regarding this type of
shock-like features, as discussed in more detail by Goodman
(2009) and Rauscher \& Menou (2010).}.

\begin{figure}
\centerline{
\includegraphics[width=3.5in,angle=0]{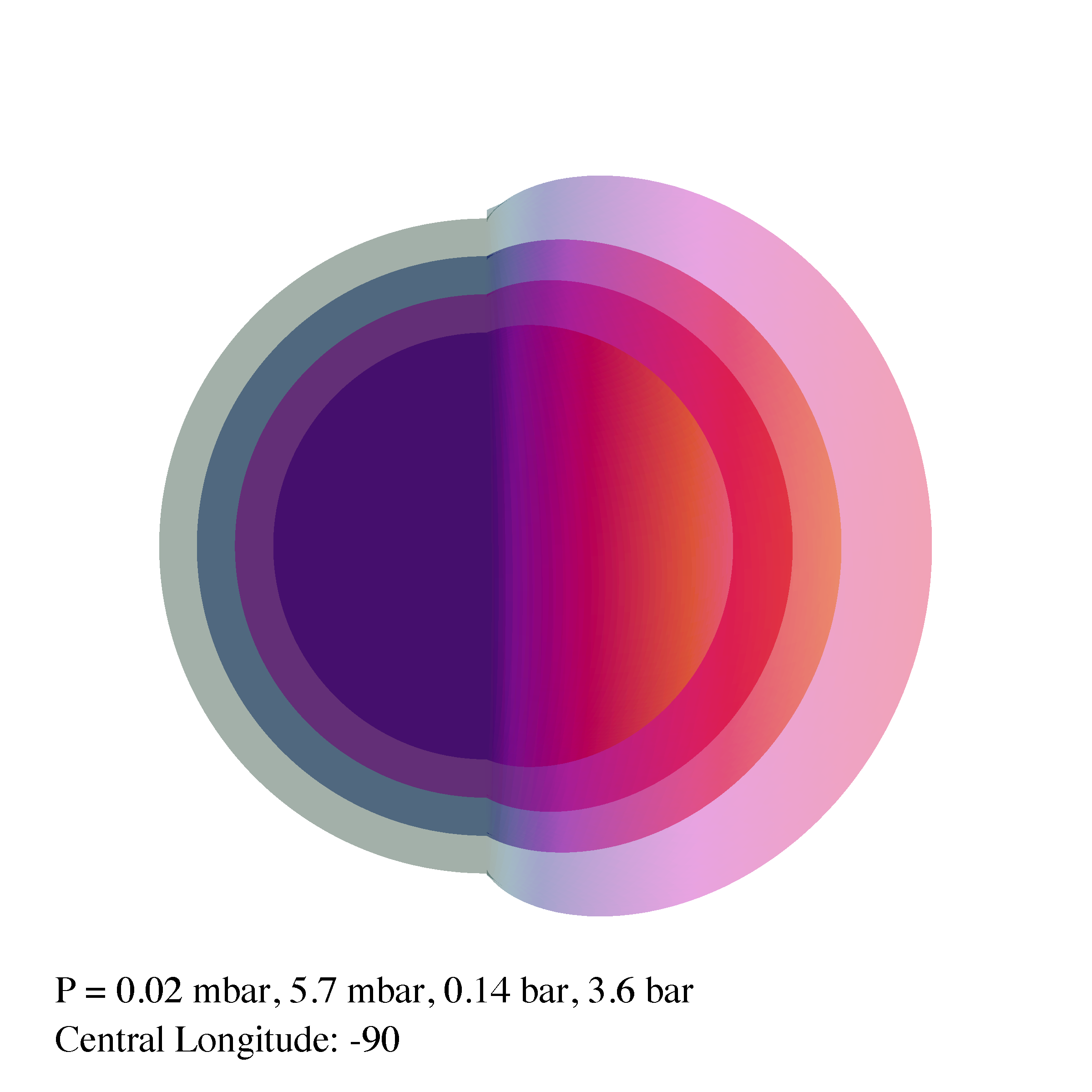}
\includegraphics[width=4.0in,angle=0]{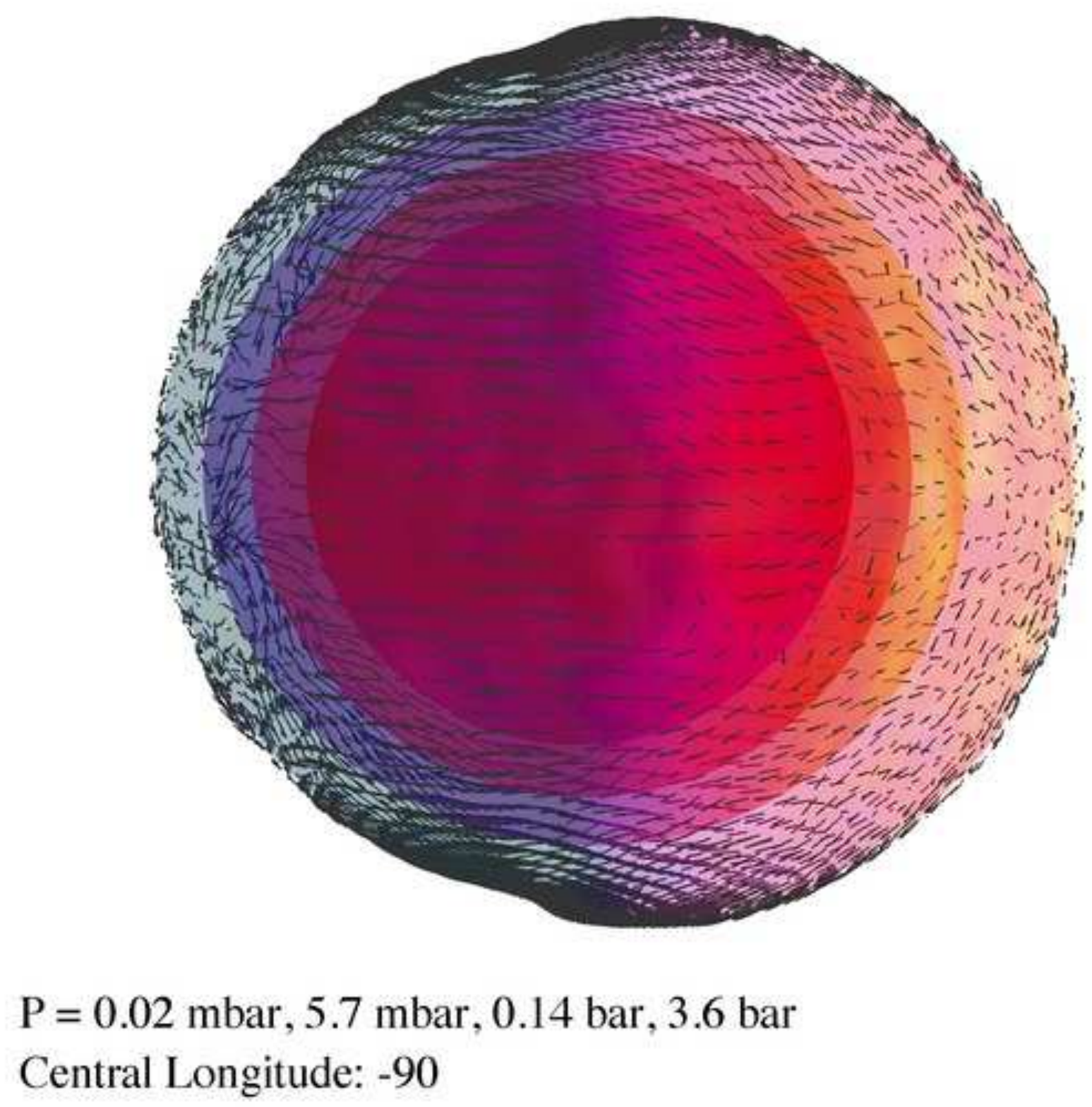}}
\vskip-1.5in
\centerline{
\includegraphics[width=3.5in,angle=0]{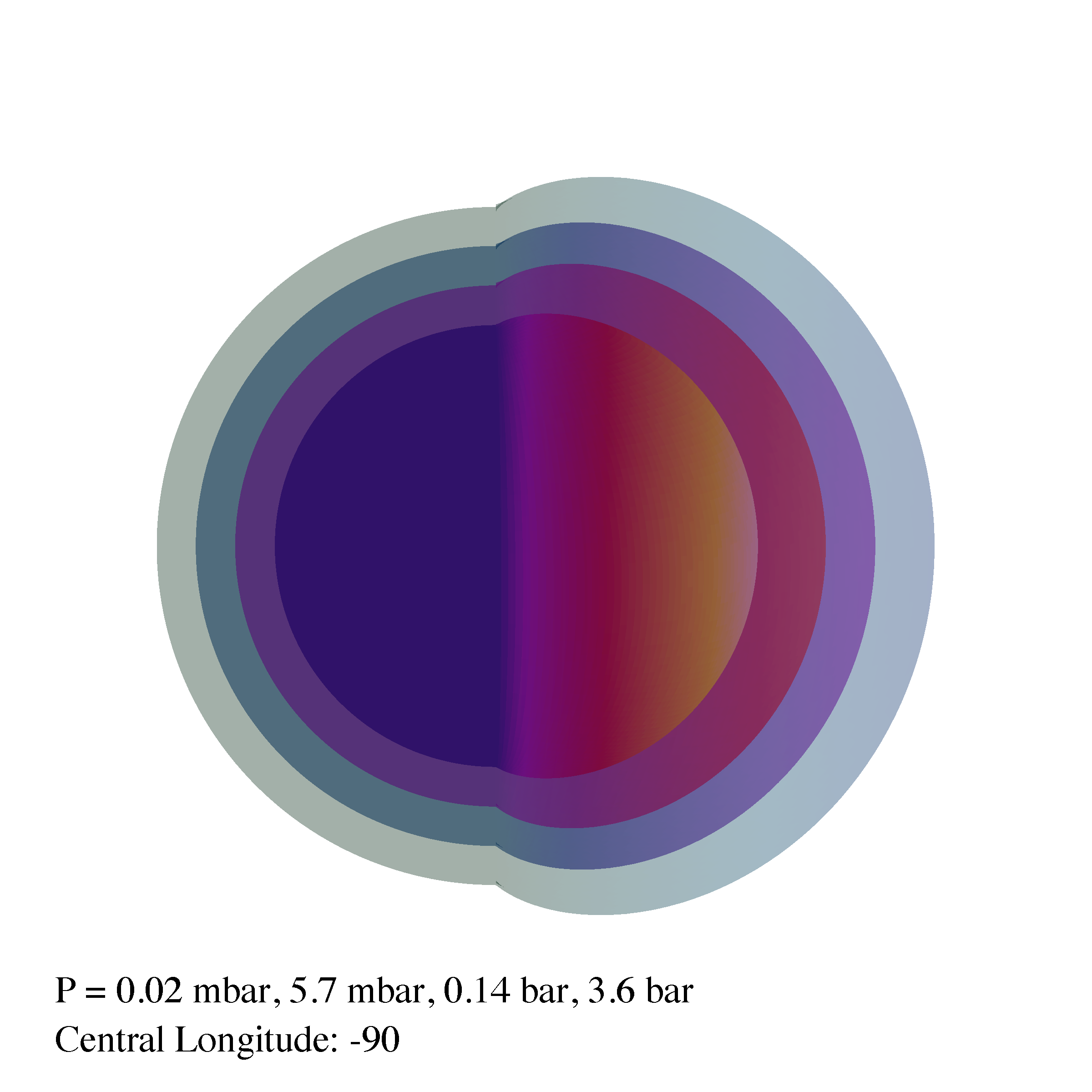}
\includegraphics[width=4.0in,angle=0]{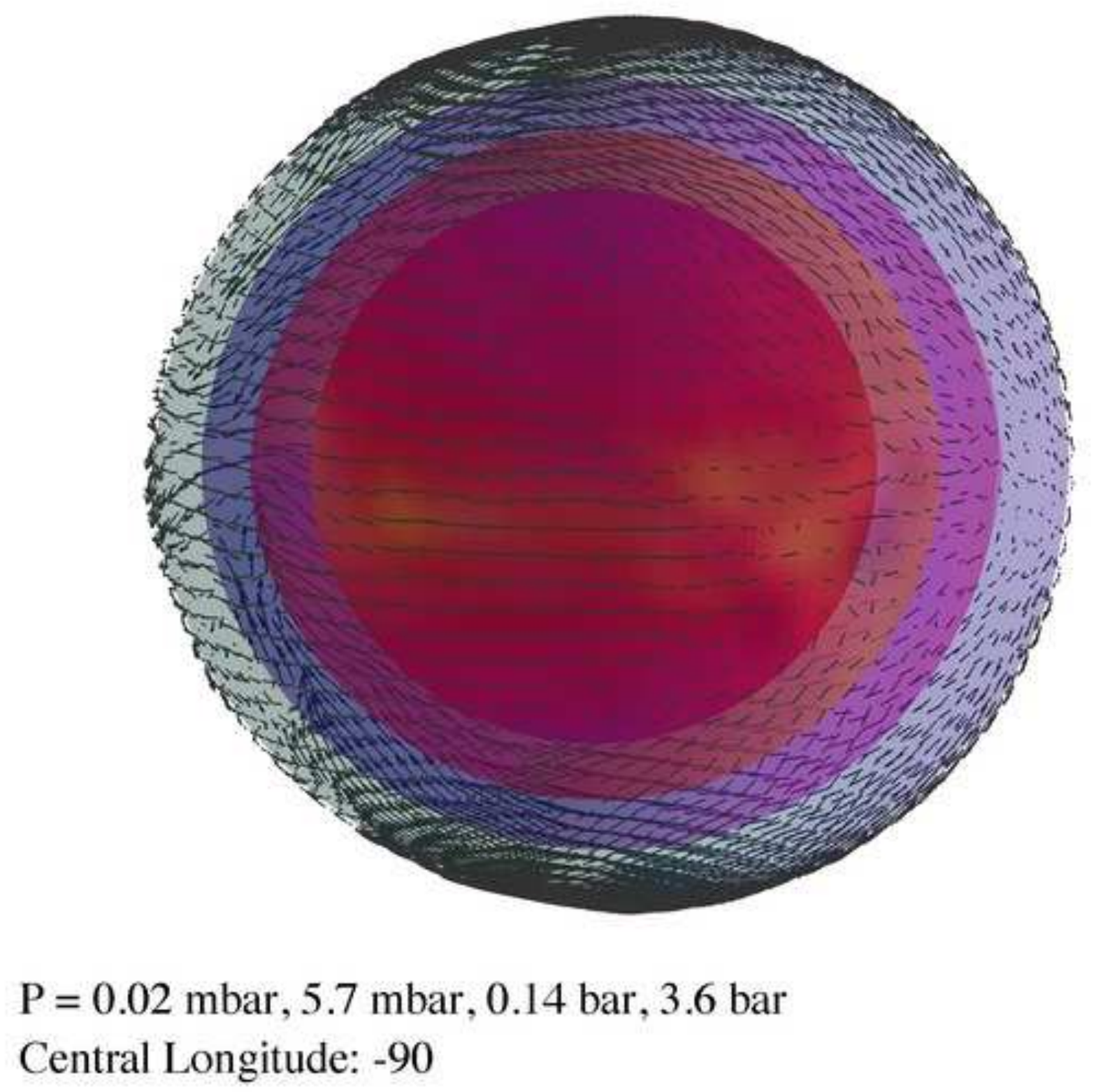}}
\caption{Temperature maps, with (right) and without (left)
dynamical redistribution.  The top panels are for the model with an extra optical absorber, and the bottom 
panels neglect such an absorber. The longitudinal flows distribute heat to the nightside, and thereby
partially even out the scale height bulge differential between the dayside and nightside. The vectors
on the right-side panels portray the local velocity fields. The vertical scales have been
exaggerated by a factor of ten to better visualize the day-night asymmetry.  Different pressure 
levels are shown, at $2.6\times 10^{-5}$, $5.7\times 10^{-3}$, $0.14$, and 3.6 bars,
and the colors depict the local temperatures on each level.  Red is hot and blue is cold.  }
\label{fig3}
\end{figure}

\begin{figure}
\includegraphics[height=.50\textheight,angle=0]{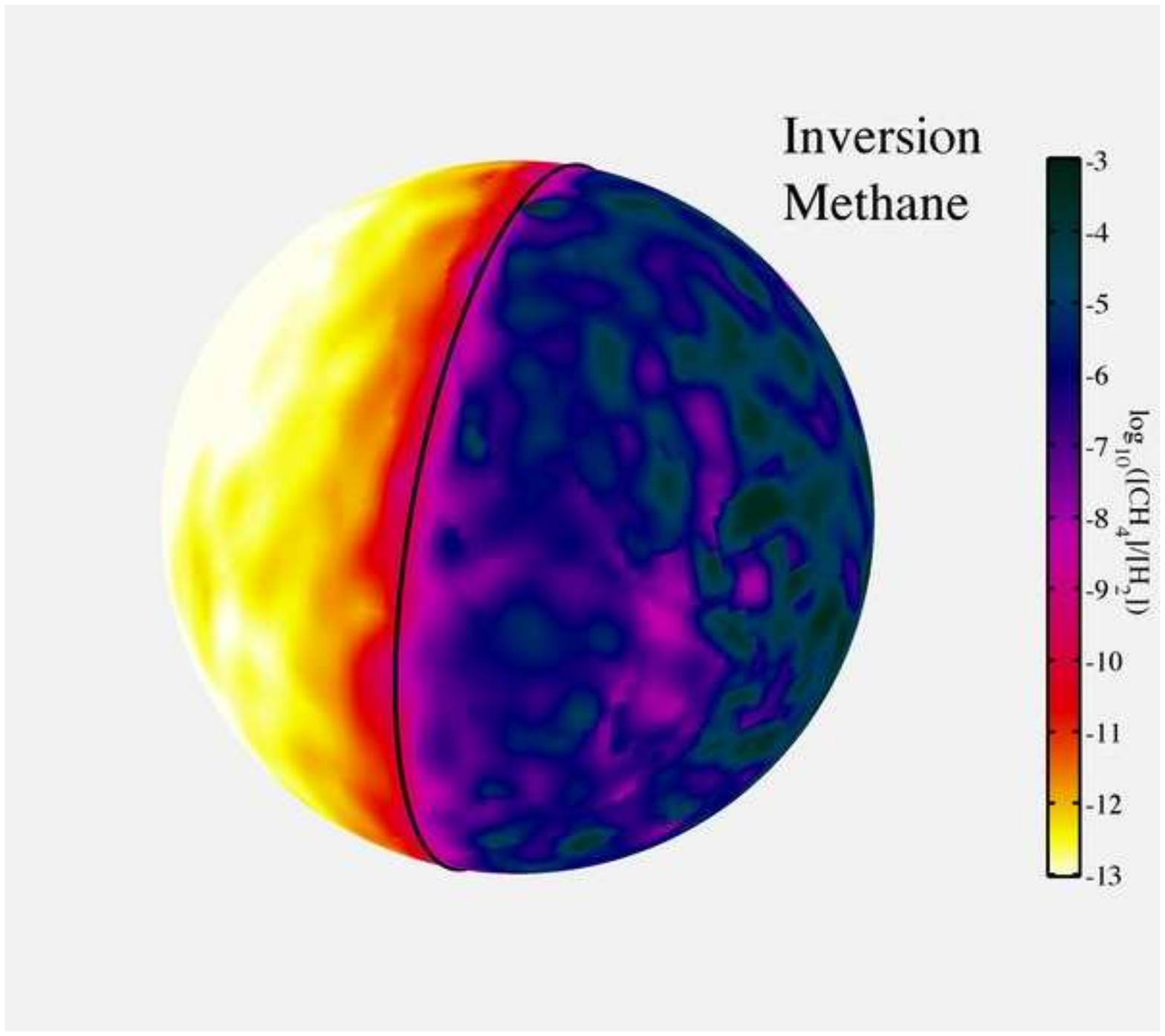}
\includegraphics[height=.50\textheight,angle=0]{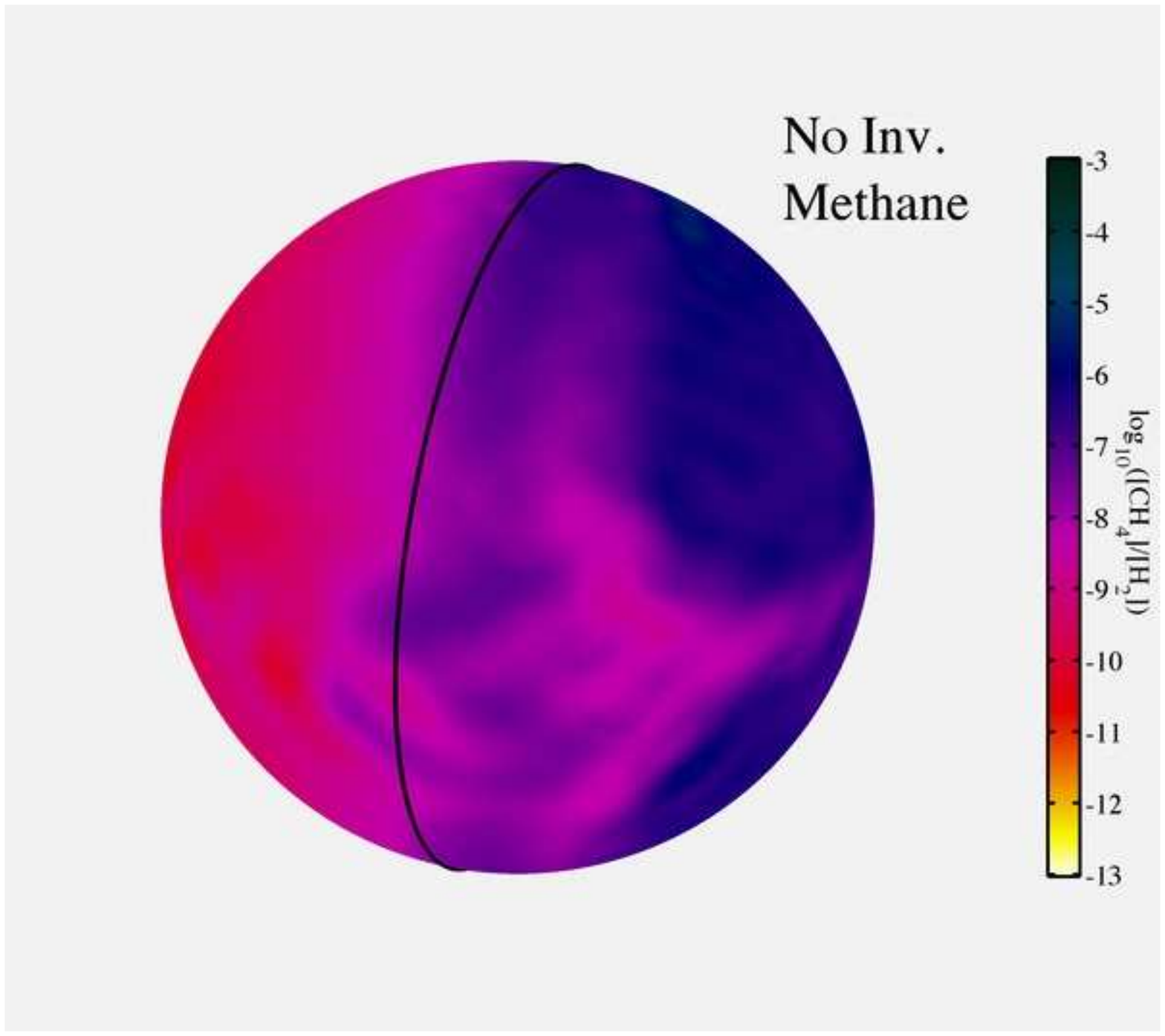}
\caption{These figures depict the distribution of methane (CH$_4$) over the surface of the 3D GCM 
models of HD 209458b we are using for this study at a pressure level of 5.7 millibars.  Chemical equilibrium 
is assumed. The left model is with and the right model is without the extra absorber.  
The nightsides for both are on the right.  Reds and yellows indicate low methane 
abundances and blues and greens represent high methane abundances.  If not in methane, 
carbon would reside in carbon monoxide in these models, which predominates on the daysides 
(particularly, in this model set, for the model with the extra optical absorber). Color bars indicate 
the logarithm (base ten) of the methane mixing ratio. The position of the terminator is clearly 
indicated with a black line. Note on the model without the extra optical absorber (right) the tongue 
of methane-depleted (CO-rich) material dragged from the dayside to the nightside in the equatorial belt.
The degree of such advection is a function of vertical height/pressure level and is greater 
for the model without the extra absorber.}
\label{fig4}
\end{figure}

Figure \ref{fig3} depicts temperature maps with and without
dynamical redistribution and with and without an extra optical absorber. 
To emphasize the contrasts, the atmospheric scale heights are exaggerated by a factor of ten.
The longitudinal flows distribute heat to the nightside, and thereby
partially even out the scale height bulge differential between the dayside and nightside. 
Additionally, Figs. \ref{fig4} depict the distribution of methane (CH$_4$) at a pressure  
level of 5.7 millibars, if in chemical equilibrium, over the surface of our 3D GCM 
models of HD 209458b, with and without an extra high-altitude optical absorber.
If not in methane, carbon would reside in carbon monoxide in these models, which predominates on the day
sides (particularly, in this model set, for the model with the extra optical absorber).  
In the model without the extra optical absorber, we note the presence of the tongue of methane-depleted      
(CO-rich) material dragged from the dayside to the nightside in the equatorial belt (see also Cooper \& Showman 2006).
The degree of such advection is a function of vertical height/pressure level.  In principle 
this feature could be detected in a precision comparison of the ingress and egress transit spectra 
(Fortney et al. 2010).  However, we find, as did Fortney et al. (2010), that this differential effect can be quite small, less than 
one percent of the already small variation with wavelength expected for the transit radius.  
Be that as it may, these two figures partially represent the model context in which we have performed 
our post-processing exercise.

\section{The Wavelength-Dependence of the Transit Radius}
\label{wave_rad}

\begin{figure}
\centerline{
\includegraphics[width=8.cm,angle=-90]{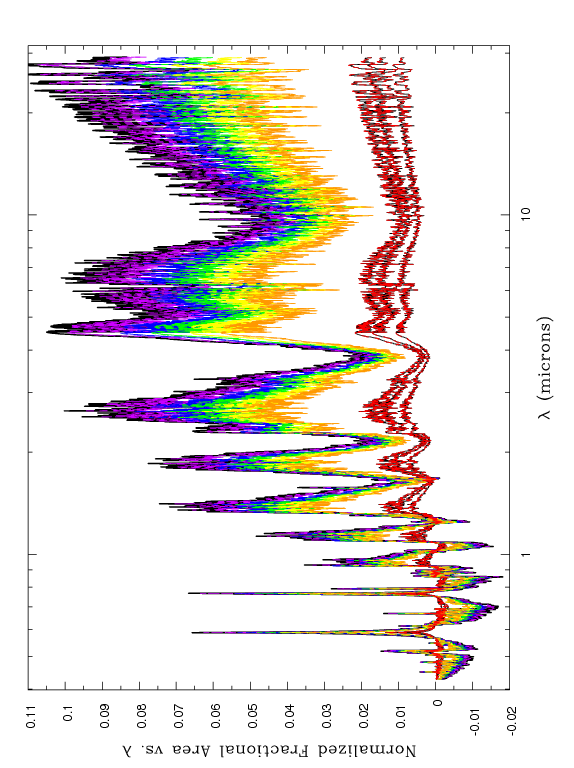}}
\centerline{
\includegraphics[width=8.cm,angle=-90]{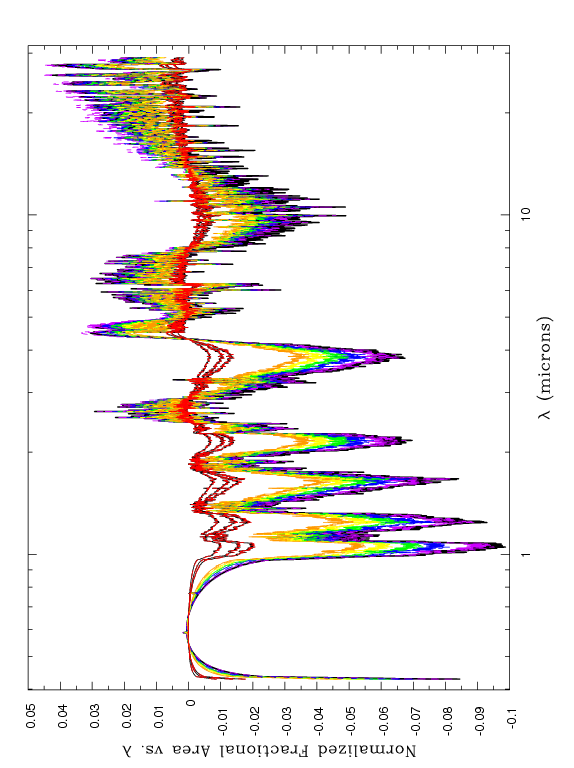}}
\caption{{\bf Top:} The normalized fractional area of the atmosphere occulting the star as a function of wavelength (in microns) from
the optical to 30 microns as the planet enters and leaves transit.  This model is without an extra 
optical absorber at altitude.  The early ingress is depicted in red and the full 
transit is the upper black curve.  The egress is also depicted, but for these models lies very near the ingress lines.
The last times of egress are depicted in black and are barely discernible under the corresponding red curves.  
The times included are -0.063, -0.062, -0.061, -0.053, -0.051, -0.048, -0.047, -0.046, -0.044, 
+0.046, +0.047, +0.048, +0.051, +0.053, +0.061, +0.062, +0.063 days relative to the time of mid-transit.  The
fractions are all normalized to the instantaneous average value in the optical wavelength range from 0.5 to 0.7 microns.
Hence, the average value of this fraction in that wavelength range (``optical") is zero.  
{\bf Bottom:} The same as the top panel, but for the model with the extra absorber at altitude. The action of the 
extra absorber is clearly seen in the optical region, and the normalization is as performed for 
the top panel.  See text for a discussion.} 
\label{fig5}
\end{figure}

The top panel of Fig. \ref{fig5} shows the normalized fractional area of 
the atmosphere occulting the star as a function of wavelength as the planet enters 
and leaves transit for the model without an extra optical absorber 
at altitude.  The fractions are all normalized to the instantaneous 
average value in the optical wavelength range from 0.5 to 0.7 microns.  Hence, 
the average value of this fraction in that wavelength range (``optical") is zero. For this model, the
base position in the atmosphere where the chord optical depth is of order unity for optical photons (which defines 
the transit radius) is deep inside the atmosphere (near 0.1 bars).  The bottom panel of Fig. \ref{fig5} 
is the same as the top panel, but for the model with the extra absorber at altitude. The action of the
extra absorber is clearly seen in the optical region, and the normalization is as performed for
the top panel.  However, for this model the chord optical depth is of order unity for a base pressure near
a millibar, much higher in the atmosphere than for the model without the extra absorber.
This difference results in perceptible differences between the two classes of models.

Water is ubiquitous in our planet model atmospheres and
its features dominate these spectra. For our models, the egress                  
and ingress are rather symmetrical, differing by no more than
$\sim$0.8\% at any wavelength, due predominantly to slight differences
in the scale heights near the respective terminators.  These small ingress/egress
differences are most prominent in the water bands, but there are slight differential
signatures in the methane and carbon monoxide bands at $\sim$1.6 \mic and
$\sim$3.3 \mic and at $\sim$2.3 \mic and $\sim$4.6 \mic, respectively, as 
might be expected from the advection of gas across the terminator depicted in Figs. \ref{fig4}.

\begin{figure}
\includegraphics[width=12.cm,angle=-90]{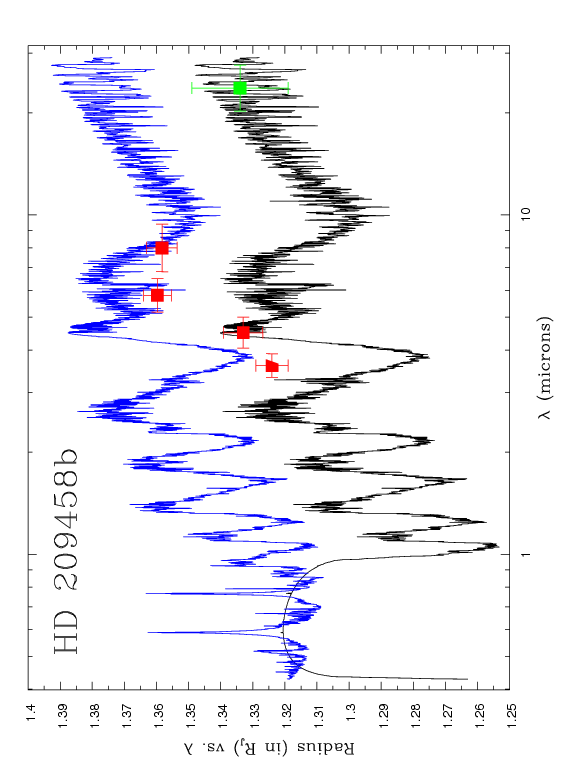}
\caption{A comparison of the transit radii during full transit as a function of wavelength (in microns) for both models 
depicted in Fig. \ref{fig3}.  No limb-darkening corrections are applied.  The normalization for both
models is to the measured radius in the optical ($\sim$1.32 \rj; Knutson et al. 2007). 
Superposed are the transit radius data in the four {\it Spitzer}/IRAC bands from Beaulieu et al. (2010)
and in the 24-micron MIPS band of {\it Spitzer} from Richardson et al. (2006). See text for a discussion.}
\label{fig6}
\end{figure}

Figure \ref{fig6} compares the transit radii during full transit as a function of wavelength 
for both models depicted in Fig. \ref{fig5}.  The normalization for both
models is to the measured radius in the optical ($\sim$1.32 \rj; Knutson et al. 2007). We believe
this to be the most physically and observationally sensible normalization. Note that when so normalized
the two models, which differ only in whether there is an extra optical absorber at altitude, predict
very different radii, though the relative differences when comparing radii at different wavelengths outside
the optical are similar. Therefore, comparing the radius as inferred from optical observations to that    
inferred from infrared observations is diagnostic of the presence of an extra absorber.  Unfortunately, the infrared transit radius data are not consistent       
with either the extra-absorber model or the no-extra-absorber model.  A comparison with the {\it Spitzer}/IRAC data from Beaulieu et al. (2010)
and the 24-micron MIPS data from Richardson et al. (2006) suggests that neither model fits both sets of data simultaneously,
but a major discrepancy is also the wide spread between IRAC 3/4 and IRAC 1/2.  The same discrepancy between model and data was
noted by Fortney et al. (2010) for their models and is as yet unexplained.

Our implementation of the extra absorber puts it throughout the upper atmosphere,
at the lowest pressures.  Given this, and as Fig. \ref{fig6} demonstrates, the corresponding
model does not evince much contrast in and out of the Na-D doublet.  This is in contradiction
with the observations of Charbonneau et al. (2002) and may hint at an upper boundary to the extent of the
optical absorber (if it exists) if the Na-D data are to be accommodated.


\section{Light Curves}
\label{phase}

The planet fluxes themselves, and their phase variation, should speak volumes
about the atmospheres of hot Jupiters, in particular their thermal and compositional 
profiles and the longitudinal dependence of their circulation regimes (Cowan \& Agol 2008).  Using the 
methodologies described in \S\ref{spectral_method}, we now present results for the 
planet-star flux ratios and photometric variations for our two models of HD 209458b 
as it traverses its orbit.

\subsection{Planet-Star Flux Ratio as a Function of Wavelength and Phase}
\label{func_wave}

\begin{figure}
\centerline{
\includegraphics[width=10.cm,angle=-90]{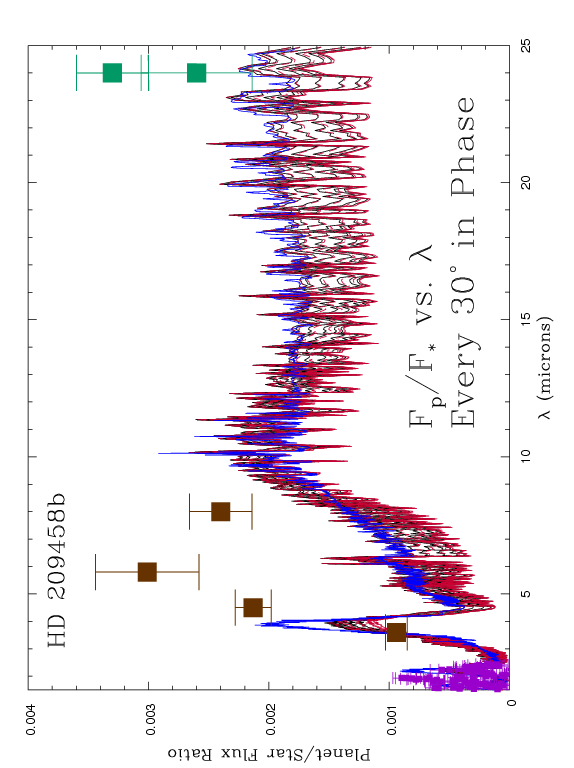}}
\centerline{
\includegraphics[width=10.cm,angle=-90]{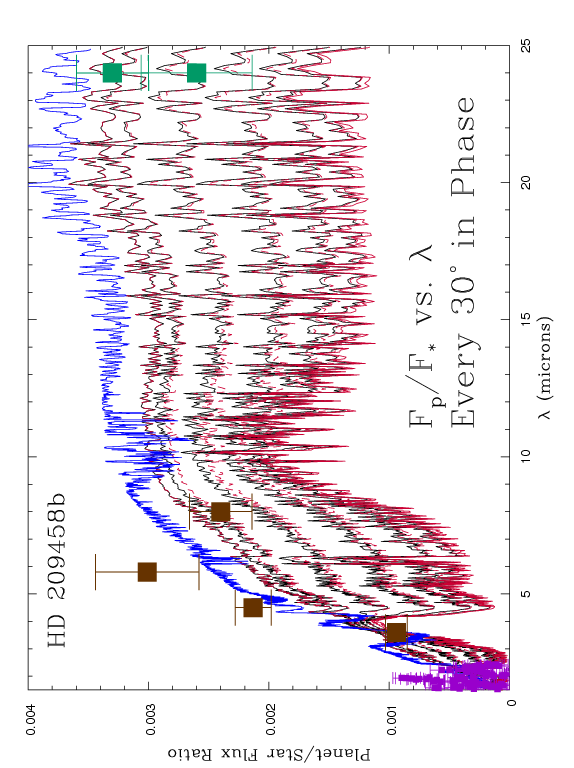}}
\caption{Planet-star flux ratios versus wavelength (in microns), for the models with (bottom) 
and without (top) the extra optical absorber, every 30$^{\circ}$ in phase of the planet's (HD 209458b's) orbit.  
Here, zero phase is at secondary eclipse. Superposed are the secondary eclipse data in 
the {\it Spitzer}/IRAC (brown) and {\it Spitzer}/MIPS 24-micron (green) bands from Knutson et al. (2008) and Deming et al. (2005), respectively.
The Deming et al. 24-micron data point is accompanied by an additional point (D. Deming, private communication). 
The $\sim$2-$\sigma$ difference between these points probably illustrates the systematic uncertainties hidden in these types of measurements.
Also included are the Swain et al. (2009) data points between 1.5 and 2.2 microns. The blue lines near the top envelope of each curve
set are the corresponding secondary eclipse predictions using the ``1D" spectral model method of Burrows et al. (2007,2008b).
The red curves are for the first 180$^{\circ}$ of the orbit and the black curves are for the second.  See text for a discussion.}
\label{fig7}
\end{figure}

Figure \ref{fig7} portrays the planet-star flux ratios versus wavelength, for the models with
and without the extra optical absorber and as a function of phase in the planet's (HD 209458b's) orbit.
Here, zero phase is at secondary eclipse. Superposed are various data sets and the corresponding
``1D" spectral models using the method of Burrows et al. (2007,2008). 
The slight differences between the zero-phase 3D calculations and the ``1D" predictions are due to the different methods
of determining the average effect of emission from the day-side hemisphere and to the dynamics inherent in 3D models.  We note many things about these curves.  First,
the data points at secondary eclipse are much more consistent with the zero-phase prediction(s) for the model with an extra absorber.
This was pointed out in Burrows et al. (2007) and Knutson et al. (2008). Second, the variation with phase from the dayside to the nightside
is much larger for the model with an extra absorber and inversion.  This is a generic feature of predictions with strong optical
absorption at altitude on the dayside. Third, there are interesting differences in the phase-dependence at different wavelengths.
The variations longward of $\sim$15 microns are generally largest.  The variation at $\sim$10 microns is muted.  The water feature near
$\sim$6.2 microns becomes relatively more prominent on the nightside, while modest on the dayside. The phase variation near $\sim$4 microns is slight.
The phase variation from $\sim$8 to $\sim$10 microns is much larger for the hot day-side model with the extra absorber. In sum, there
are important variations in infrared colors with orbital phase that may be diagnostic of the atmospheere models.

\begin{figure}
\includegraphics[height=.35\textheight,angle=0]{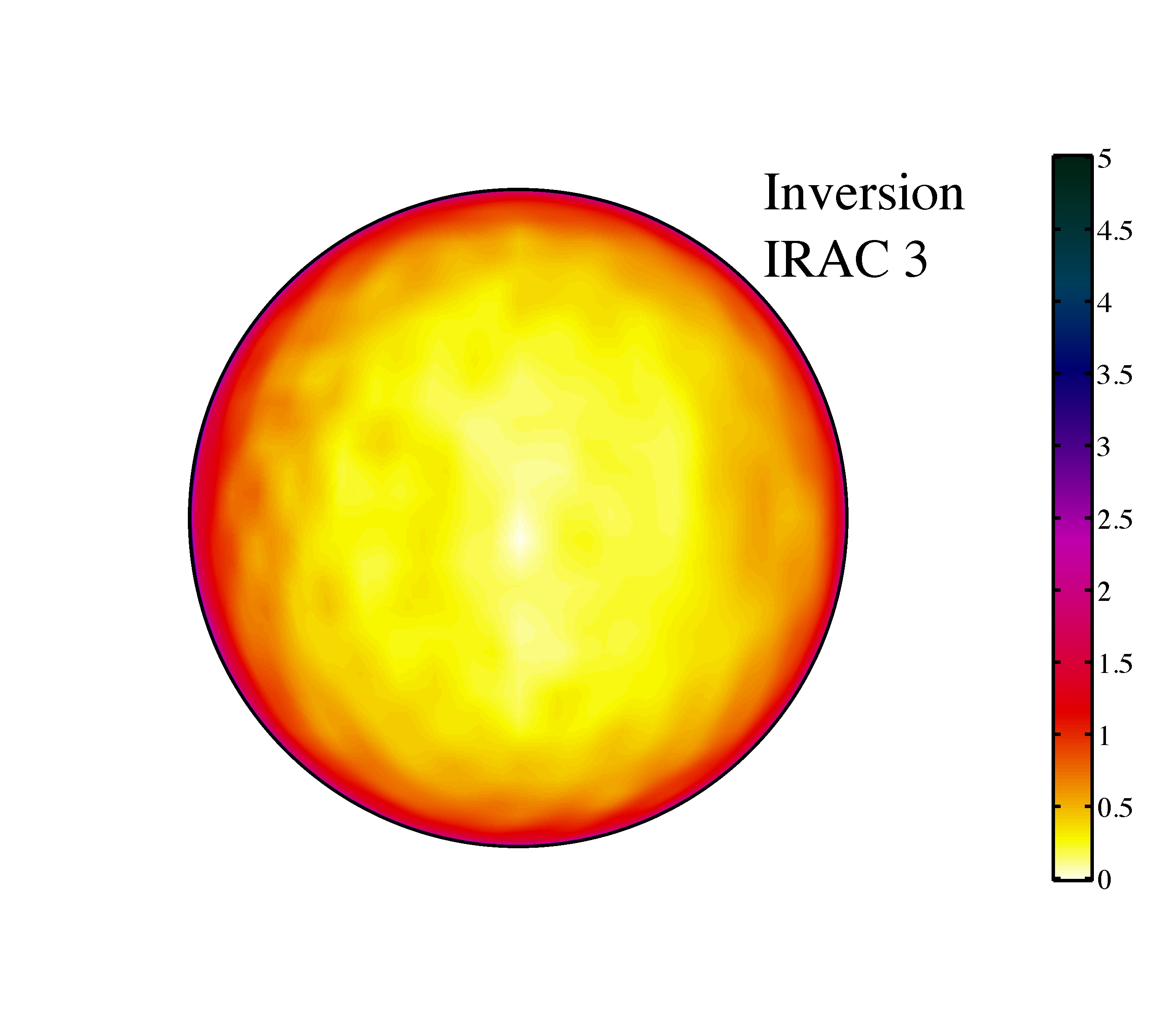}
\includegraphics[height=.35\textheight,angle=0]{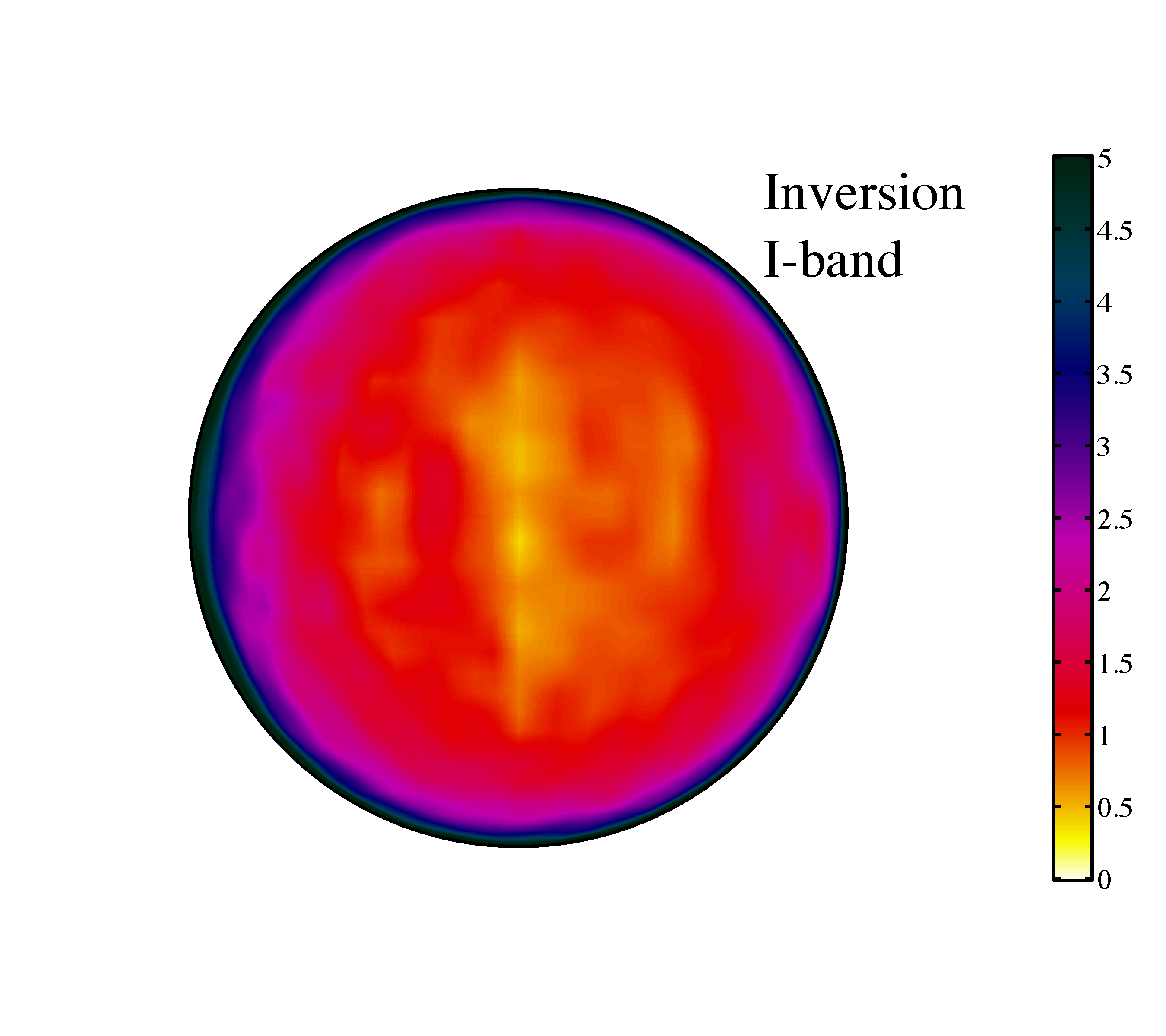}
\includegraphics[height=.35\textheight,angle=0]{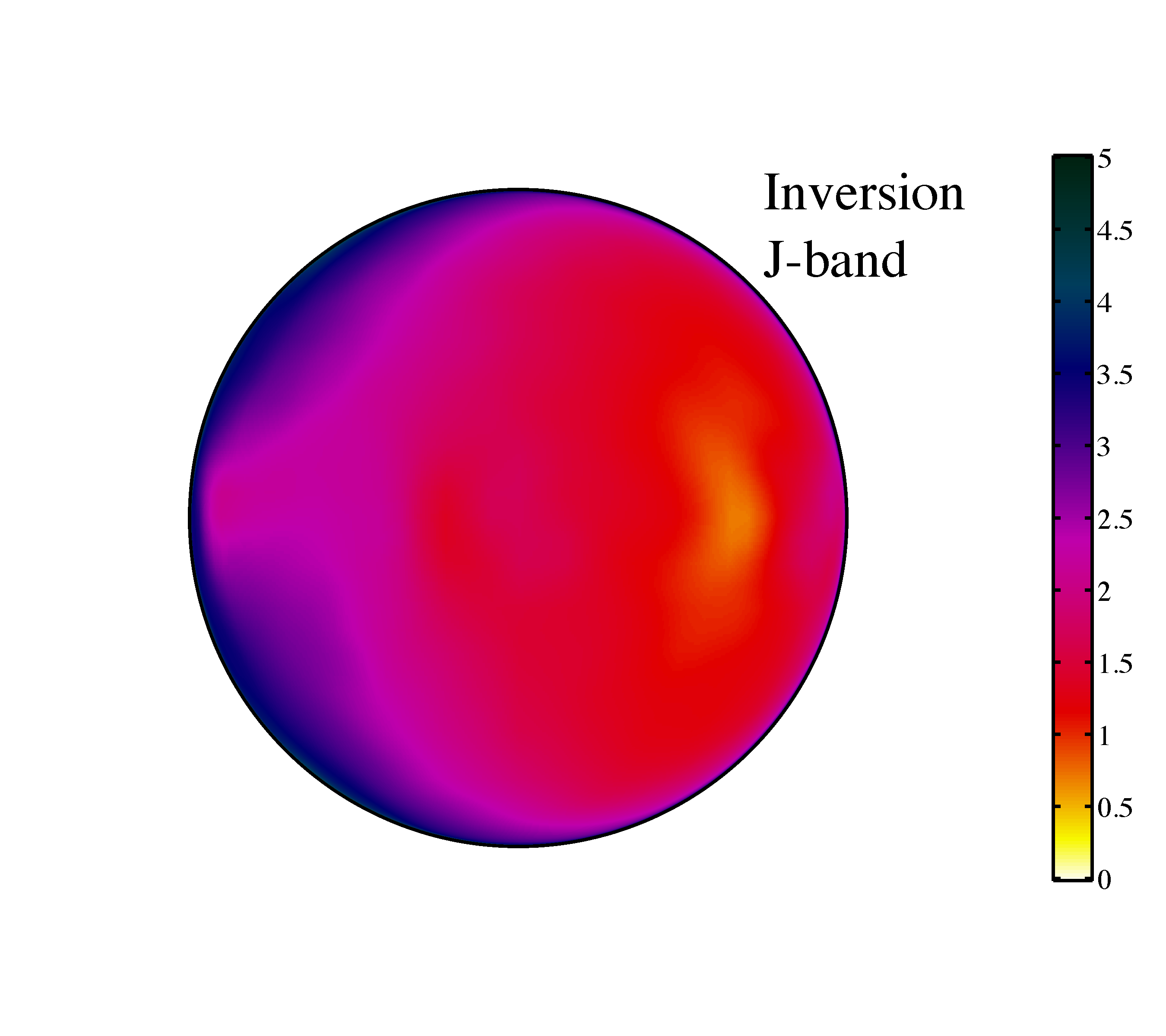}
\includegraphics[height=.35\textheight,angle=0]{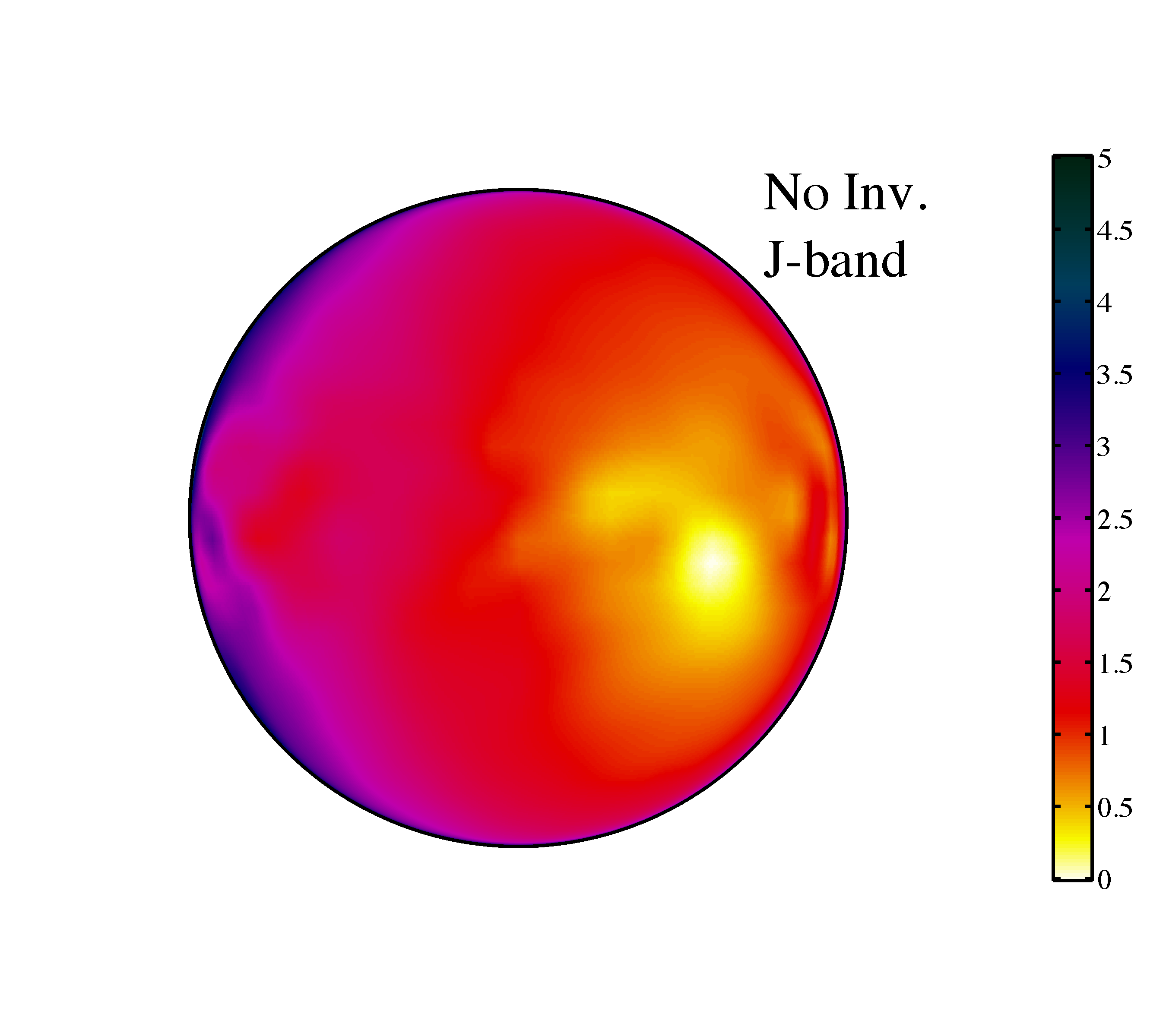}
\caption{These panels are representative planet brightness maps in various wavebands.  The top left panel is the IRAC 3
map for the model with an extra absorber; the top right panel is the $I$ band map for the same; the bottom left
panel is the $J$ band map for the model with the extra absorber; and the bottom right panel is the corresponding
$J$ band map for the model without the extra absorber.  All the maps are at full phase (secondary eclipse).
The colors and color bars are in relative magnitudes, with the brightest
regions rendered in yellow and the dimmest regions in blue.  See text for details.}
\label{fig8}
\end{figure}

The planet-star flux ratios depicted in Figs. \ref{fig7} are integrals over the surface of the planet of
brightness maps such as are shown in Figs. \ref{fig8}.  The associated panels in Fig \ref{fig8} are 
representative planet brightness maps in the IRAC 3 band, the $I$ band, and the $J$ band, the latter for the models with
and without the extra absorber. All the maps are at full phase (secondary eclipse).
In principle, these are what the planet HD 209458b would look like in ``glasses" for the various bands,
given the various relative color maps.  The color mappings depict the magnitude difference between the brightest point on the planet
and are different for each given rendering and band. Note that the hot spot in each band is 
shifted by a different angle relative to the substellar point, reflecting the different
approximate ``photospheric" pressures of the various bands and the variation in the degree and strength of zonal heat
advection with depth (Showman et al. 2009). In particular, the IRAC band fluxes, formed as they are at altitude,
are shifted the least, by no more than $\sim$20$^{\circ}$ but mostly close to $\sim$0$^{\circ}$. 
These differential phase shifts are useful diagnostics of the zonal flow regimes
and models (Cowan \& Agol 2008).

\subsection{Phase Dependence of the Photometric Band Fluxes}
\label{func_band}

\begin{figure}
\includegraphics[width=11.cm,angle=-90]{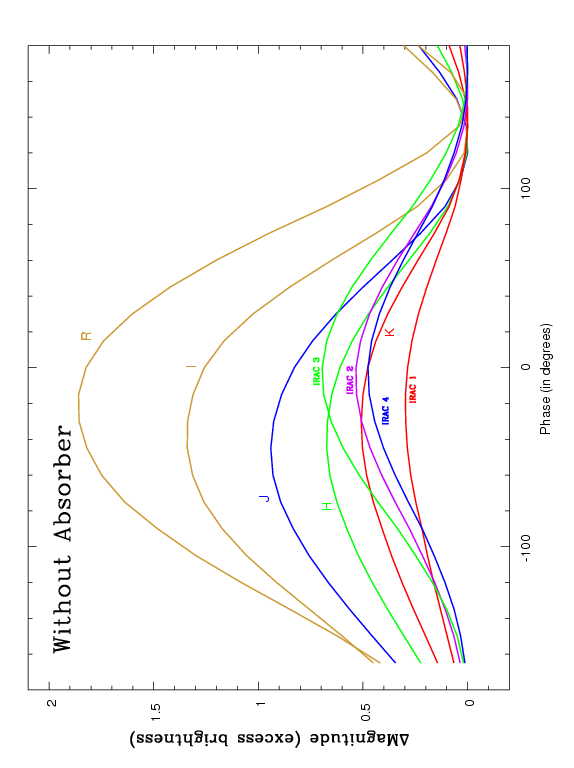}
\includegraphics[width=11.cm,angle=-90]{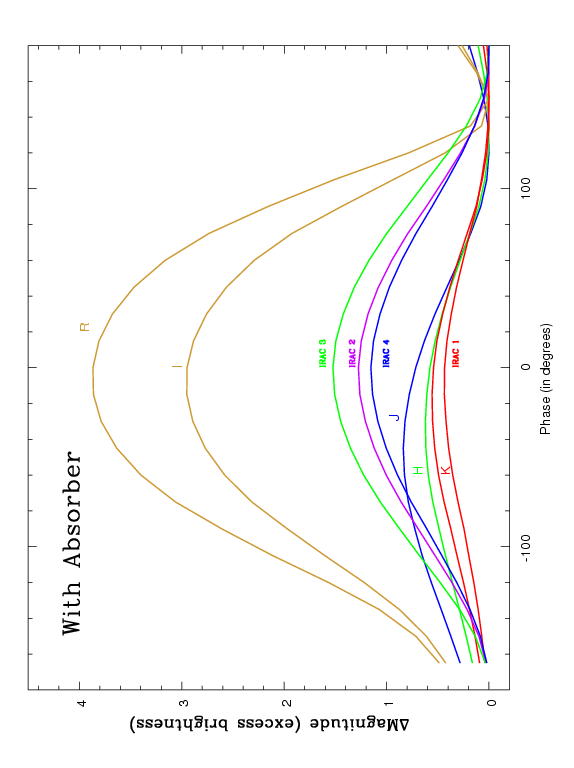}
\caption{The integrated relative light curves (in magnitudes) versus orbital phase in nine different bands ($R$, $I$, $J$, $H$, $K$, IRAC 1, IRAC 2, 
IRAC 3, and IRAC 4) for the models without (top) and with (bottom) an extra optical absorber 
at altitude and for our fiducial HD 209458b assumptions.  These data are derived from integrating 
maps such as are shown in Figs. \ref{fig6}. The magnitudes are normalized to zero at the dimmest (note !)
phases and the magnitude ranges are different for the top and bottom plots. See text for a discussion.} 
\label{fig9}
\end{figure}

This is particularly clear from Fig. \ref{fig9}, which depicts for the two different 
atmosphere models and for our fiducial HD 209458b assumptions integrated relative light 
curves versus orbital phase in the $R$, $I$, $J$, $H$, $K$, IRAC 1, IRAC 2, IRAC 3, and IRAC 4 bands.  
These data are derived by integrating maps such as are shown
in Figs. \ref{fig8}. The different phase shifts of the hot spots suggested in Figs. \ref{fig8} are manifest in these plots.
While the shifts in the IRAC bands are smallest (generally being formed at altitude at lower pressures), the shift
in IRAC 1 being the largest among these ($\sim$20$^{\circ}$ for the no-absorber model and 
$\sim$10$^{\circ}$ for the absorber model), for the model with no extra absorber the
shifts in the $I$ and $J$ bands can be $\sim$40$^{\circ}$ and $\sim$45$^{\circ}$, 
respectively. For the absorber model, the shift in the $J$ band is comparable, 
while the shift in the $I$ band is only $\sim$10$^{\circ}$.  The fact that many of the curves
are not flat on the far right of these plots near 180$^{\circ}$ is another indication  
of the skews in the phase light curves and of the (differential) dynamical
advection of heat by planetary zonal winds.  

Ignoring the IRAC bands, the contrasts are generally largest for 
bands at the shortest wavelengths.  In Fig. \ref{fig9}, the differences 
between the models with and without the high-altitude absorber are clear, with the brightness 
variations with phase being larger for the absorber model in the IRAC, $R$, and $I$ bands.  
The two models are comparable in the near-infrared bands. The specific numbers derived are not 
as important as the general trends revealed by this representative modeling exercise. 
In principle, by measuring such phase curves in different bands, one could back out
information on the thermal and wind profiles as a function of depth and longitude.  This is in part
because the different bands are formed and have approximate photospheric positions at different
pressures in the atmosphere.  Such remote sensing of a hot Jupiter atmosphere would enable a new
level of scrutiny for exoplanets and may be possible using JWST. 

\section{Conclusions}
\label{conclusions}

Using a 3D GCM with Newtonian cooling and day- and nightside equilibrium 
temperature profiles based on two 1D spectral models (with and without a thermal 
inversion and a hot upper atmosphere), we created dynamical models of 
the transiting giant exoplanet HD 209458b. We post-processed these 3D model 
atmospheres with a detailed opacity code to obtain transit radius 
spectra at various epochs during the primary transit. Then, using a spectral atmosphere code, we 
integrated over the face of the planet seen by an observer at various orbital phases
and calculated light curves as a function of wavelength and for nine different photometric 
bands from $R$ ($\sim$0.6 \mic) through IRAC 4 ($\sim$8 \mic). The products of this study are
generic predictions for the character of the expected phase variations both of a giant planet's
transit spectrum and of its light curves. Since our GCM employed Newtonian cooling (and not
radiative transfer using opacities that corresponded with those used in the post-processing), 
the calculations are slightly inconsistent.  Nevertheless, the results have interesting features
that should inform future measurements. We found that the temporal variations
in the derived integral quantities for this model suite are small (less than 1\%) and 
that the ingress/egress contrasts due to zonal flows, while also small ($< 0.5 - 1.0$ \%), are most manifest 
due to the scale-height asymmetries near the terminators and in the water bands.
While models we have generated with and without the methane and carbon monoxide bands
indicate that ingress/egress differences in the CO and CH$_4$ bands might be diagnostic 
of both carbon chemistry and abundance asymmetries due to zonal flows near the terminators,
the differential effects due to different carbon-species abundances at the different terminators 
are small (less than a few tenths of a percent).  A similar conclusion was reached by Fortney 
et al. (2010), who nevertheless found a slightly larger quantitative effect.

Since the transit radius of HD 209458b has been measured in the optical, it should not be allowed 
to differ when comparing models with and without an upper-atmosphere absorber and the predicted
radii at other wavelengths must be determined relative to it.  When this is done, the predicted radii
in the near- and mid-infrared are very different (by as much as 0.05 \rj), though our
current models do a poor job of fitting all the transit radius data for HD 209458b simultaneously
(e.g., Knutson et al. 2007; Beaulieu et al. 2010; Richardson et al. 2006). The reason for
this discrepancy is currently unknown. 

We determined that the angular phase shifts of the brightness peaks relative to the
orbital ephemeris are functions of photometric band.  The $J$, $H$, and $K$ bands are shifted most,
by as much as $\sim$45$^{\circ}$, while the IRAC bands are shifted least. This is because 
different bands are spectrally formed at different pressure depths and the zonal winds 
that advect heat are depth-dependent. The IRAC band photospheres are generally at altitude, while
those in the near infrared are at deeper pressures near and above $\sim$0.1 bars.  Therefore,
the question of the magnitude of the downwind shift in the planetary ``hot spot" due to equatorial
winds is nuanced, depending upon the waveband in which 
measurements are made.  This wavelength dependence of the phase shift in the brightness light 
curve, and the associated day/night contrasts, can in principle be used to constrain 
the circulation regimes of irradiated giant planets and to probe different pressure
levels of a hot Jupiter atmosphere.  Though our calculations focused on models of 
HD 209458b, similar calculations for other transiting hot Jupiters in low-eccentricity
orbits should yield transit spectra and light curves of a similar character.  As the 
subject of comparative exoplanetology matures, and JWST comes online, such observational 
manifestations of global circulation may well become possible for a wide range
of irradiated planets. Our models have been constructed to help prepare the way, however 
imperfectly, for that era.

\acknowledgments

We acknowledge useful conversations with Ian Dobbs-Dixon and Brad Hansen
and Ivan Hubeny for his general support of the COOLTLUSTY code. 
AB wa supported by NASA grant NNX07AG80G and under JPL/Spitzer Agreements 1328092, 1348668, and 1312647.
ER was supported by a NASA Graduate Student Research Program Fellowship, contract NNX08AT35H, and KM was supported by 
the Spitzer Space Telescope Program under contract JPLCIT 1366188.
The authors are pleased to note that part of this work was performed while 
in residence at the Kavli Institute for Theoretical Physics, funded by the NSF
through grant no. PHY05-51164.




{}

\clearpage

\clearpage
\thispagestyle{empty}

\clearpage

\clearpage

\clearpage


\setlength{\voffset}{-60mm}

\clearpage

\clearpage

\clearpage
\setlength{\voffset}{0mm}

\clearpage

\clearpage

\end{document}